\documentclass[a4paper,11pt]{amsart}
\begin{document}

\title{{\bf  Again on the non-existence \\of gravitational waves and of black holes}}
\author{Angelo Loinger}
\date{}
\address{Dipartimento di Fisica, Universit\`a di Milano, Via
Celoria, 16 - 20133 Milano (Italy)}
\email{angelo.loinger@mi.infn.it}
\thanks{In course of publication on \emph{Spacetime \& Substance}}

\begin{abstract}
Very straightforward arguments proving the physical non-existence
of GW's and of BH's. They are so simple that even the members of
the Wheelerian establishment will understand them.
\end{abstract}

\maketitle

\vskip1.20cm
\emph{\textbf{An obvious premise}}: in (the exact, i.e.
non-approximate \cite{1})  general relativity (GR) only the
concepts and the results that are independent of the choice of the
system of \emph{\textbf{general}} co-ordinates have a physical
meaning.
\par
\emph{i}) Let us consider  any solution of the Einstein field
equations of GR which has -- in a given co-ordinate system -- a
definite undulatory character, i.e. that represents a
gravitational wave (GW). Now, through a \emph{finite} sequence of
co-ordinate transformations, endowed with suitable undulating
properties, the primary wave character of our solution can be
fully \emph{obliterated}. Thus, this character has no
\emph{physical} meaning, it is only a mathematical property of the
original co-ordinate system. \emph{Q}.\emph{e}.\emph{d}. --
\par
\emph{i}') A first proof of the physical non-existence of the GW's
was given in 1917 by Tullio Levi-Civita \cite{2}. Other proofs
have been published in recent years by the present writer
\cite{3}. \par
 \emph{ii})  As it was remarked by Eddington \cite{4}, the
 solution of the problem of the Einsteinian gravitational field
 generated by a point mass $M$ (at rest) is given -- if $r$,
 $\vartheta$, $\varphi$ are spherical polar co-ordinates -- by the
 following spacetime interval:

\begin{equation} \label{eq:one}
    \textrm{d}s^{2}=\left[1-\frac{2m}{f(r)}\right]c^{2}\textrm{d}t^{2}-\left[1-\frac{2m}{f(r)}\right]
    ^{-1}\left[\textrm{d}f(r)\right]^{2}-
    \left[ f(r)\right]^{2}\left[\textrm{d}\vartheta^{2}+\sin^{2}\vartheta
    \textrm{d}\varphi^{2}\right],
\end{equation}

where $m\equiv GM/c^{2}$, and $G$ is the gravitational constant;
$f(r)$ is  \emph{any} regular function of $r$. \par Now, the
invention of the black holes (BH's) is due to a
(mis)interpretation of that form of solution (\ref{eq:one}) for
which $f(r)\equiv r$: it is the so-called \emph{standard}
solution, erroneously named ``Schwarzschild solution''. In
reality, Karl Schwarz\-schild chose
$f(r)\equiv[r^{3}+(2m)^{3}]^{1/3}$, see \cite{5}; this
\emph{original} Schwarz\-schildean form is singular only at the
spatial point $r=0$. Thus, the notion of BH does \emph{not} have a
physical meaning, it is only a (fictive) by-product of a very
particular choice of the function $f(r)$.
\emph{Q}.\emph{e}.\emph{d}. --

\normalsize \vskip0.5cm
\noindent {APPENDIX}\nopagebreak \par \nopagebreak
\vskip0.15cm\nopagebreak
 All the (erroneous)  \emph{loci communes} concerning the GW's and
 the BH's are respectively illustrated in the review articles by
 Schutz \cite{6} and by Celotti \emph{et alii} \cite{7}.

\small \vskip0.5cm\par\hfill {\emph{``Truth is aristocratic''.}
     \vskip0.10cm\par\hfill Howleglas}


\normalsize


\small
\begin{thebibliography}{99}

\bibitem{1}
Original and acute considerations about the full inadequacy of the
approximate \emph{li\-nea\-ri\-zed} version of GR to the question
of the GW's have been developed by H. Weyl, \emph{Amer. J. Math.},
\textbf{66}(1944) 591.

\bibitem{2}
T. Levi-Civita, \emph{Rend. Acc. Lincei}, \textbf{26} (1917) 381;
an English translation in \emph{arXiv:physics/9906004} (June 2nd,
1999). See also: Idem, \emph{Rend. Acc. Lincei}, \textbf{11}
(s.6a) (1930) 3 and 113.

\bibitem{3}
A. Loinger, \emph{Nuovo Cimento} B, \textbf{115} (2000) 679; Idem,
\emph{Spacetime \& Substance}, \textbf{3} (2002) 129; Idem,
\emph{ibidem} \textbf{3} (2002) 145; Idem, \emph{On Black Holes
and Gravitational Waves}, (La Goliardica Pavese, Pavia) 2002, Part
II.

\bibitem{4}
A. S. Eddington, \emph{The Mathematical Theory of Relativity},
Second Edition (Cambridge University Press, Cambridge) 1960, p.94.
See also: A. Loinger, \emph{Spacetime \& Substance}, \textbf{3}
(2002) 145; Idem, \emph{On Black Holes and Gravitational Waves},
(La Goliardica Pavese, Pavia) 2002, Part I.

\bibitem{5}
K. Schwarzschild, \emph{Berl.Ber.}, (1916) 189; an English
translation in \emph{arXiv:\-physics/9905030} (May 12th, 1999).
See also: M. Brillouin, \emph{Journ. Phys. Rad.}, \textbf{23}
(1923) 43; an English translation in \emph{arXiv:physics/0002009}
(February 3rd, 2000); K. Schwarzschild, \emph{Berl. Ber.}, (1916)
424; an English translation in \emph{arXiv:physics/9912033}
(December 16th, 1999).

\bibitem{6}
B.F.Schutz, \emph{Class. Quantum Grav.}, \textbf{16} (1999), A131;

\bibitem{7}
A. Celotti, J.C. Miller and D.W. Sciama, \emph{Class. Quantum
Grav.}, \textbf{16} (1999) A3.

\end{thebibliography}
\end{document}